\documentclass[a4paper,12pt]{article}
\usepackage{amsmath,amsfonts,amssymb,cite}

\begin{document}

\begin{center}
{\Large\bf The flavor universality of some mass splittings in
hadrons}
\end{center}

\begin{center}
{\large S. S. Afonin\footnote{E-mail: \texttt{afonin@hep.phys.spbu.ru}}}
\end{center}

\begin{center}
{\small V. A. Fock Department of Theoretical Physics, Saint-Petersburg
State University, 1 ul. Ulyanovskaya, 198504, Russia}
\end{center}

\begin{abstract}
The approximate chiral invariance of the two-flavor QCD is known
to be spontaneously broken. This effect explains the relatively
small pion mass and, as is widely believed, the mass splittings of
would-be chiral partners --- the hadrons of equal spin but
opposite parity lying in one multiplet of the chiral symmetry. We
present experimental evidences that in reality such mass
splittings in the meson sector seem to be approximately
flavor-independent in all cases where they can be tested
experimentally --- the spin 0 and 1. In addition, a partial flavor
independence holds for spin $1/2$ baryons (namely among states in
which at least one of quarks is not $u$ or $d$ one). This property
allows to predict masses and quantum numbers for 10 new hadrons.
The given flavor-independence, if confirmed for higher spins,
hints at the existence of universal scale
$\Lambda_{\text{strong}}\sim300$~MeV in the strong interactions
which, in contrast to $\Lambda_{\text{QCD}}$, is flavor and scheme
independent. Some manifestations of $\Lambda_{\text{strong}}$ are
discussed.
\end{abstract}

\section{Introduction}

One of the most challenging tasks in the physics of strong
interactions is to understand how the confinement forces generate
the observable hadron spectrum. The problem is extremely difficult
both for theoretical and experimental reasons. The theoretical
difficulty is related to a high complexity of the strong
interactions in the hadronization phase. On the other hand, the
hadron spectrum is still not well established experimentally, even
the existing data is sometimes contradictory. As a result,
accurate comparisons of model predictions with the actual data are
often problematic. The absence of complete experimental data can
be partly compensated if some general behavior of the hadron
spectrum is known in detail. This subject has attracted much
attention
recently~\cite{likhoded,cl1,cl1_b,cl1_c,cl1_d,parity,clust_rev,glozman,klempt,shif,cohen,arriola,arriola_b,wei,bugg2012,AP}
(many other references can be found in
reviews~\cite{parity,clust_rev,glozman,klempt}, some physical
implications are discussed
in~\cite{bicudo,bicudo_b,forkel,cl3,mezoir,glozman2,hydr,kirchbach,kirchbach_b,cl4,zhen}).

The strong interactions reveal two spectacular non-perturbative
phenomena --- the quark confinement and the spontaneous chiral
symmetry breaking (CSB). Both features have been inferred from
observations: The former was deduced from non-existence of free
quarks while the latter explains (among other things) small masses
of pions in comparison with typical hadron masses. In spite of a
long history of intensive researches, the relation between these
phenomena remains unclear~\cite{brambilla}. The CSB shapes in a
decisive way the hadronic world at low energies, i.e. at energies
below 1$\div$1.2 GeV~\cite{gl,gl_b}. In view of absence of
reliable analytical tools for handling the non-perturbative part
of QCD, the most confident information that we can extract about
the CSB and related physics is encoded in the actual experimental
data.

Among profound predictions of the CSB is generation of rather
large mass splitting (MS) between hadrons lying in one multiplet
of the chiral $SU_L(2)\times SU_R(2)$ group. For example, the MS
between the $\rho(770)$-meson and $a_1(1230)$-meson is usually
interpreted as a consequence of the CSB. This idea goes back to
the Weinberg sum rules~\cite{wein} and since that time it lies in
the core of many phenomenological approaches to the low-energy
strong interactions --- from extensions of the Nambu--Jona-Lasinio
model~\cite{klev} to modern bottom-up holographic models for
QCD~\cite{son1,pomarol}. The idea is that the dynamics responsible
for the appearance of (quasi)massless pions must simultaneously
lead to a considerable MS between the vector and axial-vector
mesons (in a broader context --- between the scalar and
pseudoscalar mesons, between the nuclons having the spin-parity
$J^P=\frac12^+$ and the $N(1535)$-nucleon having $J^P=\frac12^-$
etc.).

The main purpose of the present note is to draw attention to one
remarkable property of the experimental hadron spectrum. It
consists in a surprising universality of mass splittings between
hadrons of opposite parity at very different energy scales.

The quark masses represent input parameters in the QCD Lagrangian.
The masses of $u$- and $d$-quarks are of the order of several
MeV~\cite{pdg} that is much less than the typical hadronic scale
(1 GeV). This results in the approximate $SU_L(2)\times SU_R(2)$
chiral symmetry of the strong interactions relating particles of
opposite parity. The hadrons made of $u$- and $d$-quarks should
fill multiplets of the chiral group (see, e.g., the
review~\cite{glozman}). The CSB entails large MS between the
parity partners belonging to the same multiplet. Imagine now that
we enlarge the value of one or both quark masses to the GeV
scales. The approximate chiral symmetry does not exist any more.
What happens with MS of the former chiral partners? Actually the
answer is provided by the known data on heavy mesons and baryons.
This data does not indicate any significant dependence of the MS
under consideration on the quark masses. The approximate flavor
independence of MS between parity partners is puzzling and seems
to encode an important piece of information about the strong
interactions.

\section{Analysis of experimental data}

The relevant experimental data on mesons taken from the Particle
Data~\cite{pdg} is displayed in Table~1. Various MSs are shown in
Table~2. Consider the MS $M(1^{++})-M(1^{--})$. In the light
non-strange mesons (the $qq$-states in Table~1), the corresponding
isovector vector and axial-vector mesons fill the multiplet of
representation $(0,1)+(1,0)$ of the $SU_L(2)\times SU_R(2)$ chiral
group. Enlarging the quark masses, {\it apriori} we expect a
substantial change of this MS since the underlying dynamics
governing the MS is not chiral any more. In reality, we seem to
observe oscillation of the MS near 414~MeV.

\begin{table}[h]
\caption{\small The masses of ground meson states taken for our
analysis~\cite{pdg}. The experimental errors (in brackets) are
given with the accuracy of 1~MeV. The quark content is shown (it
is implied that antiquark stays instead of one of quarks). The
symbol $q$ denotes the $u$- or $d$-quark. The charged component is
considered for mesons with open flavor and different charges of
quarks. For definiteness, we take the isovector $qq$-mesons except
the scalar case (the chiral symmetry relates the isoscalar mesons
with the pions). The assignment of $ss$-states is not very
reliable: We confront the $\eta$-meson to $J^{PC}=0^{-+}$ state
(motivated by the dominance of the strange component in its mass),
$f_1(1420)$ to $J^{PC}=1^{++}$ and $f'(1520)$ to $J^{PC}=2^{++}$.}
\vspace{-0.6cm}
\begin{center}
{\tiny
\begin{tabular}{|c|cccccccccc|}
\hline
$M(J^{PC})$ & $qq$ & $qs$ & $ss$ & $qc$ & $sc$ & $cc$ & $qb$ & $sb$ & $cb$ & $bb$\\
\hline
$M(0^{-+})$ & $140$ & $494$ & $548$ & $1870$ & $1969(1)$ & $2984(1)$ & $5279$ & $5367$ & $6275(2)$ & $9398(3)$\\
%\hline
$M(1^{--})$ & $776$ & $892$ & $1020$ & $2010$ & ---  & $3097$ & $5325$ & $5416(2)$ & --- & $9460$\\
%\hline
$M(1^{++})$ & 1230(40) & $1272(7)$ & $1426(1)$ & $2423(2)$ & $2460$  & $3511$ & $5723(2)$ & $5829$ & --- & $9893$\\
%\hline
$M(0^{++})$ & $475(75)$ & --- & $980(20)$ & $2319(29)$ & $2318(1)$  & $3414$ & --- & --- & --- & $9859$\\
%\hline
$M(2^{++})$ & $1318$ & $1426(2)$ & $1525(5)$ & $2462(1)$ & $2464(2)$  & $3556$ & $5743(5)$ & $5840$ & --- & $9912$\\
\hline
\end{tabular}}
\end{center}
\end{table}
\vspace{-1cm}
\begin{table}[h]
\caption{\small Some mass splittings based on the data from
Table~1. The mean values for approximately universal mass
splittings are shown.}
\vspace{-0.6cm}
\begin{center}
{\tiny
\begin{tabular}{|c|cccccccccc|c|}
\hline
$\Delta M(J^{PC})$ & $qq$ & $qs$ & $ss$ & $qc$ & $sc$ & $cc$ & $qb$ & $sb$ & $cb$ & $bb$ & Mean\\
\hline
$M(1^{++})-M(1^{--})$ & $454(40)$ & $380(7)$ & --- & $413(2)$ & ---  & $414$ & $398(2)$ & $413(2)$ & --- & $433$ & $414(22)$\\
%\hline
$M(0^{++})-M(0^{-+})$ & $335(75)$ & --- & $432(20)$ & $449(29)$ & $349(1)$ & $430(1)$ & --- & --- & --- & $461(3)$ & $409(54)$\\
\hline
$M(1^{--})-M(0^{-+})$ & $636$ & $398$ & --- & $141$ & ---  & $113(1)$ & $46$ & $46(2)$ & --- & $62(3)$ & \\
%\hline
$M(1^{++})-M(0^{++})$ & $755(85)$ & --- & $446(20)$ & $110(29)$ & $142(1)$ & $97$ & --- & --- & --- & $34$ & \\
%\hline
$M(2^{++})-M(1^{++})$ & $78(40)$ & $154(7)$ & $99(5)$ & $32(2)$ & $4(2)$ & $45$ & $20(5)$ & $11$ & --- & $19$ & \\
\hline
\end{tabular}}
\end{center}
\end{table}

If we consider the spontaneously broken $U_L(1)\times U_R(1)$
symmetry, the relevant multiplet is given by the pair
$(\omega,f_1)$ with MS about 500~MeV. Thus, qualitatively we are
dealing with the same phenomenon.

In the scalar sector, the situation is very similar. The ground
isoscalar scalar and isovector pseudoscalar $qq$-states form the
representation $\left(\frac12,\frac12\right)$ of the chiral group.
The data in the heavy quark sector is scarcer, nevertheless they
happen to show a very similar effect. Moreover, the averaged MS
turns out to be remarkably close to the vector one. It is not
excluded that they coincide within the experimental errors.

We also provide in Table~2 some specific MSs between states with
equal angular momentum $L$ of quark-antiquark pair. According to
the standard quark model, the $S$-wave quark-antiquark state
($L=0$) can have the quantum numbers of vector or pseudoscalar
particle while the $P$-wave one ($L=1$) can be realized as scalar,
axial-vector or spin-two tensor resonance of positive parity. It
is seen that the MSs $M(1^{--})-M(0^{-+})$ ($S$-wave) and
$M(1^{++})-M(0^{++})$, $M(2^{++})-M(1^{++})$ ($P$-wave) degrease
rapidly in response to increasing the quark masses.

The effect of approximate universality of MS between equal-spin
parity partners also persists in the baryon sector, at least for
the spin $1/2$ baryons. The available data on the ground baryon
states is given in Table~3~\cite{pdg}. The corresponding MSs
within the parity partners are presented in Table~4. It is clearly
seen that if one or more $q$-quarks are replaced by heavier $s$,
$c$ or $b$ quarks, the MS is immediately reduced by a factor of 2
and becomes a surprisingly stable quantity with the mean value
$295\pm8$~MeV. For higher spin baryons, there is only one MS where
this effect can be tested --- the spin $3/2$ baryons (see
Table~4). As we see, the test is passed.

\begin{table}[h]
\caption{\small The masses of baryons relevant for our
analysis~\cite{pdg}. The names are indicated in square brackets.}
\vspace{-0.1cm}
\begin{center}
{\tiny
\begin{tabular}{|c|cccccc|}
\hline
$M(J^P)$ & $qqq$ & $qqs$ & $qss$ & $sss$ & $qqc$ & $qqb$\\
\hline
$M\left(\frac12^+\right)$ & $938$ $[N]$ & $1116$ $[\Lambda]$ & $1322$ $[\Xi]$ & --- & $2286$ $[\Lambda]$ & $5619$ $[\Lambda]$ \\
%\hline
$M\left(\frac12^-\right)$ & $1535(10)$ $[N]$ & $1405(1)$ $[\Lambda]$ & --- & --- & $2592$ $[\Lambda]$ & $5912$ $[\Lambda]$ \\
%\hline
$M\left(\frac32^+\right)$ & $1232(2)$ $[\Delta]$ & --- & $1532$ $[\Xi]$ & $1672$ $[\Omega]$ & --- & --- \\
%\hline
$M\left(\frac32^-\right)$ & $1700(30)$ $[\Delta]$ & --- & $1823(1)$ $[\Xi]$ & --- & --- & --- \\
\hline
\end{tabular}}
\end{center}
\end{table}
\vspace{-1cm}
\begin{table}[h]
\caption{\small The mass splittings for baryons from Table~3. The
mean value is cited for all states except the ground $N$ and
$\Delta$.} \vspace{-0.1cm}
\begin{center}
{\tiny
\begin{tabular}{|c|cccccc|c|}
\hline
$\Delta M(J^P)$ & $qqq$ & $qqs$ & $qss$ & $sss$ & $qqc$ & $qqb$ & Mean\\
\hline
$M\left(\frac12^-\right)-M\left(\frac12^+\right)$ & $597(10)$ & $281(1)$ & --- & --- & $306$ & $293$ & $295 \pm 8$\\
%\hline
$M\left(\frac32^-\right)-M\left(\frac32^+\right)$ & $468(30)$ & --- & $291(1)$ & --- & --- & --- & \\
\hline
\end{tabular}}
\end{center}
\end{table}

\section{Predictions}

The universality of MS between the would-be chiral partners and
the approximate rate of degreasing the MS in states possessing
identical $L$ allow to predict the masses of all unknown mesons in
Table~1 and of some unknown baryons in Table~3. Below we provide
the ensuing predictions.

The known masses $M(0^{-+})$ for the $qs$, $qb$, $sb$, and $cb$
states and the universal MS $M(0^{++})-M(0^{-+})=409\pm54$~MeV
give new scalar mesons
$$
M_{sq}(0^{++})=903\pm54~\text{MeV},
$$
$$
M_{qb}(0^{++})=5688\pm54~\text{MeV},
$$
$$
M_{sb}(0^{++})=5766\pm54~\text{MeV},
$$
$$
M_{cb}(0^{++})=6684\pm54~\text{MeV}.
$$

From the existence of $M_{sc}(1^{++})$ and approximately universal
MS $M(1^{++})-M(1^{--})=414\pm22$~MeV we conjecture a new vector
meson
$$
M_{sc}(1^{--})=2046\pm22~\text{MeV}.
$$

We can estimate the mass $M_{cb}(1^{--})$ from the rate of
convergence of MS $M(1^{--})-M(0^{-+})$. Namely, from the known
value of this MS in the $cb$ and $bb$ sector (see Table~2) we may
estimate its value in the $cb$ sector as $55\pm10$~MeV. Then
$M_{cb}(1^{--})$ follows from $M_{cb}(0^{-+})$,
$$
M_{cb}(1^{--})=6330\pm10~\text{MeV}.
$$
The MS $M(1^{++})-M(1^{--})$ yields
$$
M_{cb}(1^{++})=6744\pm24~\text{MeV}.
$$
The MS $M(2^{++})-M(1^{++})$ in the $cb$ sector is estimated from
the $sb$ and $bb$ ones as $15\pm5$~MeV. This leads to the
prediction
$$
M_{cb}(2^{++})=6759\pm25~\text{MeV}.
$$
It must be of course understood that the last three predictions
have a lower confidence level.

The averaged MS of baryon parity partners for the states
containing at least one $s$, $c$ or $b$ quark is $295\pm 8$~MeV.
Then from Tables~3 and~4 we have the predictions
$$
[\Xi]:\quad M_{qss}\left(\frac12^-\right)=1617\pm8~\text{MeV},
$$
$$
[\Omega]:\quad M_{sss}\left(\frac32^-\right)=1967\pm8~\text{MeV}.
$$

Thus, we see that a simple observation of certain regularities in
the hadron spectrum discussed in the previous Section provides
interesting estimates for masses of 10 new hadrons.

\section{Discussions}

Our short phenomenological analysis was motivated by the ongoing
discussions in the literature concerning details of global
behavior of the hadron spectrum and extraction of physical
information from the established
regularities~\cite{likhoded,cl1,parity,clust_rev,glozman,klempt,shif,cohen,arriola,wei,bugg2012,AP,bicudo,forkel,cl3,mezoir,glozman2,hydr,kirchbach,cl4,zhen}.
One of important questions can be formulated as follows: Does the
hadron mass represent a single valued function (with the accuracy,
say, several pro cent) of the total spin $J$ or of the relative
angular momentum of quarks $L$ (having fixed the other quantum
numbers and the quark masses)? In the sector of light hadrons, it
is often advocated that $J$ is the correct choice because $L$
represents a non-relativistic quantity while the light hadrons are
ultrarelativistic systems~\cite{glozman}. This point of view is
partly supported by some hadron string models. Nevertheless, the
light mesons are most naturally systematized in terms of
$L$~\cite{phen,bugg}. In addition, the Regge behavior $M^2\sim
L+n$ was observed~\cite{clust_rev,shif}, here $n$ denotes the
radial quantum number and $L$ fits well the quantum-mechanical
prediction $L=J,J\pm S$, where $S=0,1$ is the sum of spins of
quark and antiquark.

In our case, we are dealing with the ground states, $n=0$. The
assumption about the Regge behavior is not needed. The MS between
states with equal $L$ but different $J$ diminishes by an order of
magnitude when going from the light mesons to the heavy ones (see
Table~2). This behavior is natural because the heavy quarkonia
represent non-relativistic systems in which $L$ defines the total
energy containing a relatively small contribution from the
spin-orbital interactions. Consider the MS $\Delta=M(L=1)-M(L=0)$
for the heavy mesons, where $L$ is well defined. The
phenomenological property $\Delta>0$ is intuitively clear since
higher $L$ is related to larger separation, hence, to stronger
interactions between quarks. A remarkable feature of $\Delta$
consists in the fact that it seems to depend weakly both on the
quark masses and on the total spin, as we have demonstrated in
Section~2. Even more remarkably looks the fact that the given
property approximately persists in the light mesons. This
phenomenon strongly suggests that the underlying dynamics
responsible for the value of $\Delta$ is more or less universal
for any flavor.

The data on baryons is more puzzling. As follows from Table~4,
only a half of MS between $N(1535)$ and the proton has a universal
flavor-independent origin. The universal MS in mesons (for spin 0
and 1), $\Delta_{\text{M}}\sim 400$~MeV is larger than the
universal MS in baryons (for spin $1/2$ and spin $3/2$),
$\Delta_{\text{B}}\sim 300$~MeV (here we exclude the baryons
composed of light $u$ and $d$ quarks
--- $N$ and $\Delta$). The fraction
$\Delta_{\text{M}}/\Delta_{\text{B}}\sim4/3$ is remarkably close
to the Casimir factor $C_F$ of $SU(3)$ gauge group,
$C_F=(N_c^2-1)/(2N_c)=4/3$ for $N_c=3$, which represents the color
factor associated with gluon emission from a quark.

In summary, the experimental data on hadrons seems to suggest that
the strong interactions possess a universal non-perturbative
flavor-independent scale $\Lambda_{\text{strong}}$. This scale
should not be confused with $\Lambda_{\text{QCD}}$. We recall that
$\Lambda_{\text{QCD}}^{(n_f)}$ is inferred from the perturbation
theory and depends both on the exploited scheme of calculations
and on the number of quark flavors $n_f$. For instance, in the
$\overline{\text{MS}}$ scheme,
$\Lambda_{\text{QCD}}^{(3)}=340\pm8$~MeV,
$\Lambda_{\text{QCD}}^{(4)}=297\pm8$~MeV,
$\Lambda_{\text{QCD}}^{(5)}=214\pm7$~MeV~\cite{pdg}. The universal
MS in baryons $\Delta_{\text{B}}$ can be taken as a definition of
$\Lambda_{\text{strong}}$. Then the data indicates on the relation
$\Delta_{\text{M}}\simeq C_F\Delta_{\text{B}}$. It would be very
interesting to derive this relation theoretically, at least in
some limit. In fact, the characteristic scale
$\Lambda_{\text{strong}}$ is parametrically contained in the
low-energy chiral~\cite{rujula} and potential quark models in the
form of constituent quark mass $M_{\text{con}}\sim300$~MeV. Our
aim, in essence, was to emphasize the universality of this scale
for any energy region where the hadron resonances are observed.

\section{Conclusions}

The mass splitting in hadron parity pairs has, most likely, a
universal flavor-independent origin in mesons and partly universal
origin in baryons. This is directly seen in the spin 0 and 1
mesons and, to a certain extend, in the spin $1/2$ and (less
evident) $3/2$ baryons. The experimental discovery of 10 new
states described in Section~3 would be a spectacular confirmation
for our hypothesis. The observation of the same property in the
higher spin hadrons would give another confirmation. The effect
could be also tested in the lattice simulations.

The given flavor-independence, if confirmed, means that, in the
hadronization regime, the strong interactions between quarks
always contain a low-energy contribution that depends strongly on
the parity of a hadron state and weakly on its spin and quark
masses. This very part precludes the linear realization of chiral
symmetry in the light hadrons and is indispensable to
understanding the spectroscopy of the heavy hadrons. The
derivation of fraction
$\Delta_{\text{M}}/\Delta_{\text{B}}\sim4/3$ is a challenge for
dynamical models of the strong interactions.

\section*{Acknowledgments}

The author acknowledges Saint-Petersburg State University for a research grant
11.38.189.2014. The work was also partially supported by the
RFBR grant 13-02-00127-a.


\begin{thebibliography}{99}
\bibitem{likhoded} S.~S.~Gershtein, A.~K.~Likhoded and
A.~V.~Luchinsky, Phys. Rev. D {\bf 74}, 016002 (2006).
\bibitem{cl1} S.~S.~Afonin, Phys. Lett. B {\bf 639}, 258 (2006).
%%CITATION = PHLTA,B639,258;%%
\bibitem{cl1_b} S.~S.~Afonin, Eur. Phys. J. A {\bf 29}, 327
(2006).
%%CITATION = HEP-PH/0606310;%%
\bibitem{cl1_c} S.~S.~Afonin, Phys. Rev. C {\bf 76}, 015202
(2007).
%%CITATION = PHRVA,C76,015202;%%
\bibitem{cl1_d} S.~S.~Afonin, Mod. Phys. Lett. A {\bf 23}, 3159 (2008).
%%CITATION = MPLAE,A23,3159;%%
\bibitem{parity} S.~S.~Afonin, Int. J. Mod. Phys. A {\bf 22}, 4537 (2007).
%%CITATION = ARXIV:0704.1639;%%
\bibitem{clust_rev} S.~S.~Afonin, Mod. Phys. Lett. A {\bf 22}, 1359 (2007).
%%CITATION = MPLAE,A22,1359;%%
\bibitem{glozman} L.~Ya.~Glozman, Phys. Rept. {\bf 444}, 1 (2007).
\bibitem{klempt} E. Klempt and A. Zaitsev, Phys. Rept. {\bf 454}, 1 (2007).
\bibitem{shif} M. Shifman and A. Vainshtein, Phys. Rev. D {\bf 77}, 034002 (2008).
\bibitem{cohen} T.~D.~Cohen, Nucl. Phys. Proc. Suppl. {\bf 195}, 59 (2009).
\bibitem{arriola} P.~Masjuan, E.~Ruiz Arriola and W.~Broniowski,
Phys. Rev. D {\bf 85}, 094006 (2012).
\bibitem{arriola_b} P.~Masjuan, E.~Ruiz Arriola and W.~Broniowski,
Phys. Rev. D  {\bf 87}, 118502 (2013).
\bibitem{wei} K.~-W.~Wei and X.~-H.~Guo, Phys. Rev. D {\bf 81}, 076005 (2010).
\bibitem{bugg2012} D.~V.~Bugg, Phys. Rev. D {\bf 87}, 118501 (2013).
\bibitem{AP} S.~S.~Afonin and I.~V.~Pusenkov, arXiv:1308.6540 [hep-ph].
%%CITATION = ARXIV:1308.6540;%%
\bibitem{bicudo} P. Bicudo, Phys. Rev. D {\bf 76}, 094005 (2007).
\bibitem{bicudo_b} P. Bicudo, Phys. Rev. D {\bf 81}, 014011 (2010).
\bibitem{forkel} H.~Forkel, M.~Beyer and T.~Frederico, JHEP {\bf 0707}, 077 (2007).
\bibitem{cl3} S.~S.~Afonin, Nucl. Phys. B {\bf 779}, 13 (2007).
%%CITATION = HEP-PH/0606291;%%
\bibitem{mezoir} E.~H.~Mezoir and P.~Gonzalez, Phys. Rev. Lett. {\bf 101}, 232001 (2008).
\bibitem{glozman2} L.~Ya.~Glozman, arXiv:0811.0470 [hep-ph].
\bibitem{hydr} S.~S.~Afonin, Int. J. Mod. Phys. A {\bf 23}, 4205 (2008).
\bibitem{kirchbach} M.~Kirchbach, AIP Conf. Proc. {\bf 1488}, 236
(2012).
\bibitem{kirchbach_b} M.~Kirchbach and C.~B.~Compean, AIP Conf. Proc. {\bf 1361}, 208 (2011).
\bibitem{cl4} S.~S.~Afonin and A.~D.~Katanaeva, Eur. Phys. J. C {\bf 73}, 2529 (2013).
%%CITATION = ARXIV:1307.6936;%%
\bibitem{zhen} Q.~Zhen, D.~Xin-Ping and W.~Ke-Wei, Chin. Phys. C {\bf 37}, 053102 (2013).
\bibitem{brambilla} N. Brambilla {\it et al.},  arXiv:1404.3723 [hep-ph].
\bibitem{gl} J. Gasser and H. Leutwyler, Ann. Phys. (NY) {\bf 158}, 142
(1984).
\bibitem{gl_b} J. Gasser and H. Leutwyler, Nucl. Phys. B~{\bf 250}, 465 (1985).
\bibitem{wein} S. Weinberg, Phys. Rev. Lett. {\bf 18}, 507 (1967).
\bibitem{klev} S. P. Klevansky, Rev. Mod. Phys. {\bf 64}, 649 (1992).
\bibitem{son1} J.~Erlich, E.~Katz, D.~T.~Son and M.~A.~Stephanov,
Phys. Rev. Lett. {\bf 95}, 261602 (2005).
\bibitem{pomarol} L.~Da Rold and A.~Pomarol, Nucl. Phys. B {\bf 721}, 79 (2005).
\bibitem{pdg} J. Beringer {\it et al.} (Particle Data Group), Phys. Rev. D {\bf 86}, 010001 (2012).
\bibitem{phen} A.~V.~Anisovich, V.~V.~Anisovich and A.~V.~Sarantsev,
Phys. Rev. D~{\bf 62}, 051502(R) (2000).
\bibitem{bugg} D.~V.~Bugg, Phys. Rept. {\bf 397}, 257 (2004).
\bibitem{rujula} A. De Rujula, H. Georgi and S. L. Glashow, Phys. Rev. D {\bf 12}, 147
(1975).
%\bibitem{halzen} F. Halzen, A. D. Martin, {\it Quarks and Leptons} (John Wiley \& Sons, 1984).
%\bibitem{graz} C. B. Lang and M. Schr\"{o}ck, Phys. Rev. D {\bf 84}, 087704 (2011).
\end{thebibliography}
\end{document}